\documentclass[pre,superscriptaddress,12pt]{revtex4-1}
\usepackage{graphicx}
\usepackage{latexsym}
\usepackage{amsmath}
\usepackage{bm}
\newcommand{\beq}{\begin{equation}}
\newcommand{\eeq}{\end{equation}}
\newcommand{\ba}{\begin{array}{ccc}}
\newcommand{\ea}{\end{array}}
\newcommand{\nn}{\nonumber \\}

\newcommand{\br}{{\bm r}}
\newcommand{\bk}{{\bm k}}

\newcommand{\bQ}{{\bm Q}}
\newcommand{\bvec}[1]{\boldsymbol{#1}}
\def\bea{\begin{eqnarray}}
\def\eea{\end{eqnarray}}
\begin{document}
\title{Comment on\\ ``Symmetry classification of bond order parameters in cuprates''}

\author{Andrea Allais}
\affiliation{Department of Physics, Harvard University, Cambridge MA 02138}

\author{Johannes Bauer}
\affiliation{Department of Physics, Harvard University, Cambridge MA 02138}

\author{Subir Sachdev}
\affiliation{Department of Physics, Harvard University, Cambridge MA
02138}
 \affiliation{Perimeter Institute for Theoretical Physics, Waterloo, Ontario N2L 2Y5, Canada}

\date{\today}

\begin{abstract}
We review the transformation of bond order waves with non-trivial form factors under time-reversal and point group symmetry.
Zeyher (arXiv:1406.6846) argues that certain $d$-form factor states must be ``flux states'',
but this does not apply to the form factors as defined by us (arXiv:1402.4807). The latter definitions were used in 
the experimental detection (arXiv:1402.5415, arXiv:1404.0362).
\end{abstract}

\maketitle

\section{Introduction}

A recent STM experiment \cite{seamus} has presented sublattice-resolved information 
on the density wave order in the underdoped cuprates,  including a direct phase-sensitive identification of a
$d$-form factor. X-ray experiments have also reported evidence for such a form-factor \cite{comin2}.

In this note we reiterate and clarify symmetry aspects of such form factors, and specifically their transformation properties under time-reversal and point group symmetry. It is useful to define the generalized bilinear order parameter $\Delta_{ij}$ by \cite{max,rolando,jay,allais1,allais2,pepin,chubukov,DCSS,SWSS}
\bea
\Delta_{ij} &\equiv& \left\langle c_{i \alpha}^\dagger c_{j \alpha}^{\vphantom\dagger} \right\rangle \nn
&=&
\sum_{\bQ} \left[ \frac{1}{V} \sum_{\bk} e^{i \bk \cdot (\br_i - \br_j)} \Delta_\bQ (\bk) \right] e^{i \bQ \cdot (\br_i + \br_j)/2} ,
\label{Dij}
\eea
with $i,j$ labels for the Cu sites of the square lattice at spatial co-ordinates $\br_i$, $\br_j$, and $c_{i \alpha}$ is the electron annihilation
operator with spin label $\alpha = \uparrow, \downarrow$.
The wavevectors of the density wave orders are $\bQ$, and $\Delta_\bQ (\bk)$ are the corresponding form factors which are complex
functions of the 
wavevector $\bk$ extending over the first Brillouin zone. The volume of the system is $V$. In momentum space, we can write
Eq.~(\ref{Dij}) as
\beq
\Delta_\bQ (\bk) = \left\langle c_{\bk-\bQ/2,\alpha}^\dagger c_{\bk+\bQ/2,\alpha}^{\vphantom\dagger} \right\rangle.
\label{Dij2}
\eeq
Taking the Hermitian conjugate of Eq.~(\ref{Dij}), we note that every such order parameter must satisfy
\beq
\Delta_{ij} = \Delta_{ji}^\ast \label{h1}
\eeq
which immediately implies that
\beq
\Delta_{\bQ}^\ast (\bk)  = \Delta_{-\bQ} (\bk). \label{h2}
\eeq
Then, we note that under time-reversal, $\mathcal{T}$, we have
\beq
\mathcal{T} : \quad \Delta_{ij} \rightarrow \Delta_{ij}^\ast . \label{t1}
\eeq
The combination of Eqs.~(\ref{Dij},\ref{h1},\ref{h2},\ref{t1}) is now seen to imply
\beq
\mathcal{T} : \quad \Delta_\bQ (\bk) \rightarrow \Delta_{\bQ} (-\bk) = \Delta_{-\bQ}^\ast (- \bk). \label{t2}
\eeq
It is this simple time-reversal transformation which is the main advantage of the parameterization in Eq.~(\ref{Dij}).

In contrast, numerous other works \cite{chetan,zeyherflux,sudip,vojta4,dunghai,jhu}, including the recent work of Zeyher \cite{zeyher}, use a 
parmetrization of the form
\begin{equation}
\Delta_{ij} 
= \sum_{{\bf Q}}  \left[ \frac{1}{V} \sum_{{\bk}} F_\bQ ( {\bk}) e^{i {\bk} \cdot \left( {\br}_i
- {\br}_j \right)} \right] e^{i {\bQ} \cdot  {\bf r}_i}, \label{e2}
\end{equation}
or equivalently
\beq
F_\bQ (\bk) = \left\langle c_{\bk,\alpha}^\dagger c_{\bk+\bQ,\alpha}^{\vphantom\dagger} \right\rangle.
\eeq
Comparing Eqs.~(\ref{Dij}) and (\ref{e2}) we can conclude that
\beq
F_\bQ ({\bk} - \bQ/2) = \Delta_\bQ ({\bk}). \label{FD}
\eeq
Thus the ``$d$-wave flux'' state \cite{zeyherflux,sudip}, which has $\bQ = (\pi, \pi)$ and $F_\bQ (\bk) \sim \cos (k_x) - \cos(k_y)$, is a $p$-form factor state in our notation, with $\Delta_\bQ (\bk) \sim \sin (k_x) - \sin(k_y)$.
Also, note that now
\beq
F_\bQ^\ast (\bk) = F_{-\bQ} (\bk + \bQ),
\eeq
and under time-reversal
\beq
\mathcal{T} : \quad F_\bQ (\bk) \rightarrow F_{\bQ} (-\bk - \bQ) = F_{-\bQ}^\ast (- \bk) . \label{t3}
\eeq
From Eq.~(\ref{t3}) we see that a $d$-form factor \cite{vojta4,dunghai,jhu}
\beq
F_\bQ (\bk) \sim \cos (k_x) - \cos (k_y)
\eeq
is not invariant
under time-reversal for general $\bQ$, as has been noted by Zeyher \cite{zeyher}.
However, precisely for this reason, we have consistently used \cite{max,rolando,jay,seamus,allais1,allais2} 
Eq.~(\ref{Dij}) rather than Eq.~(\ref{e2}): the $d$-form factor
\beq
\Delta_\bQ (\bk) \sim \cos (k_x) - \cos (k_y) \label{Dq} 
\eeq
is indeed invariant under time-reversal for all $\bQ$, as is evident
from Eq.~(\ref{t2}).
Note that the $\mathcal{T}$-preserving form factor in Eq.~(\ref{Dq}), when expressed in terms of $F_\bQ (\bk)$ using
Eq.~(\ref{FD}), yields a function $F_\bQ (\bk)$ which does {\em not\/} transform under an irreducible representation of the
point group, but is a mixture of $d$- and $p$-form factors. 

\section{Point group symmetry}

Zeyher \cite{zeyher} classifies the functions $F_\bQ (\bk)$ in terms of 
irreducible point-group representations, but this does not commute with time-reversal.
Here, we consider a point-group classification of $\Delta_\bQ (\bk)$ and show by
explicit construction that
\begin{itemize}
 \item time reversal invariant, $d$-form factor bond order waves form 
 bases for \emph{irreducible} representations of the point group of the square lattice,
 \item some of these representations contain unidirectional (single wavevector) waves.
\end{itemize}

This is important because of the following argument. In a second order phase transition that breaks a symmetry group $G$, close to the critical point in the symmetry broken phase, the order parameter $\Phi$ is very small, and hence transforms under a \emph{linear} representation $\Gamma$ of $G$:
\begin{equation}
 g\in G:\quad \Phi_i \mapsto \Gamma(g)_{ij} \Phi_j\,.
\end{equation} 

If we further assume that $\Gamma$ is unitary, then it must also be \emph{irreducible}. In fact, we can always decompose the representation $\Gamma$ in irreducible representations, and, expanding the free energy to quadratic order, we have
\begin{equation}
 F = F_0 + \sum_{r,s}[\Phi^{(r)}_i]^\star c^{rs}_{ij} \Phi^{(s)}_j\,,
\end{equation} 
where $\Phi^{(r)}$ transforms under an irreducible representation $\Gamma_r$ of $G$. Since the free energy must be invariant under $G$, by Shur's lemma we have
\begin{align}
c^{rs}_{ij} = c_r \delta_{ij}\delta_{rs}\,, && F = F_0 + \sum_r c_r[ \Phi^{(r)}_i]^\star \Phi^{(r)}_i\,.
\end{align} 
For a generic, non fine-tuned phase transition, only one of the $c_r$ changes sign, and hence the order parameter transforms under the irreducible representation $\Gamma_r$.

To be self contained, let us briefly summarize properties of the symmetry group of the square $C_4$. The group is generated by $R$, a rotation of $\pi/2$ and $P$, the reflection about the $y$ axis. It contains 8 elements
\begin{equation}
 C_4 = \{1,\, R,\, R^2,\, R^3,\, P,\, RP,\, R^2P,\, R^3P\}\,.
\end{equation} 

It has 4 inequivalent irreducible representations:
\begin{enumerate}
\item The faithful, 2 dimensional representations $\Gamma_{\mathrm{f}}$
\begin{align}\label{faithful}
&\Gamma_{\mathrm{f}}(R)  = \begin{pmatrix}0 & -1\\1 &  0\end{pmatrix}\,,
&&\Gamma_{\mathrm{f}}(P) = \begin{pmatrix}-1 &  0\\0 & 1\end{pmatrix}\,,
\end{align} 
which is just a subgroup of $O(2)$.
\item A one-dimensional representation $\Gamma_{\mathrm{p}}$ in which the normal subgroup $\{1,\,R,\, R^2,\, R^3\}$ is mapped to the identity
\begin{align}
 &\Gamma_{\mathrm{p}}(R) = 1\,, &&\Gamma_{\mathrm{p}}(P) = -1\,.&
\end{align} 
\item A one-dimensional representation $\Gamma_{\mathrm{r}}$ in which the normal subgroup $\{1,\,R^2,\, P,\, R^2P\}$ is mapped to the identity
\begin{align}
 &\Gamma_{\mathrm{r}}(R) = -1\,, &&\Gamma_{\mathrm{r}}(P) = 1\,.&
\end{align} 
\item The trivial, one-dimensional representation $\Gamma_{0}(P) = \Gamma_0 (R) = 1$.
\end{enumerate}

\begin{figure}[ht]
\begin{center}
\includegraphics[scale=0.5]{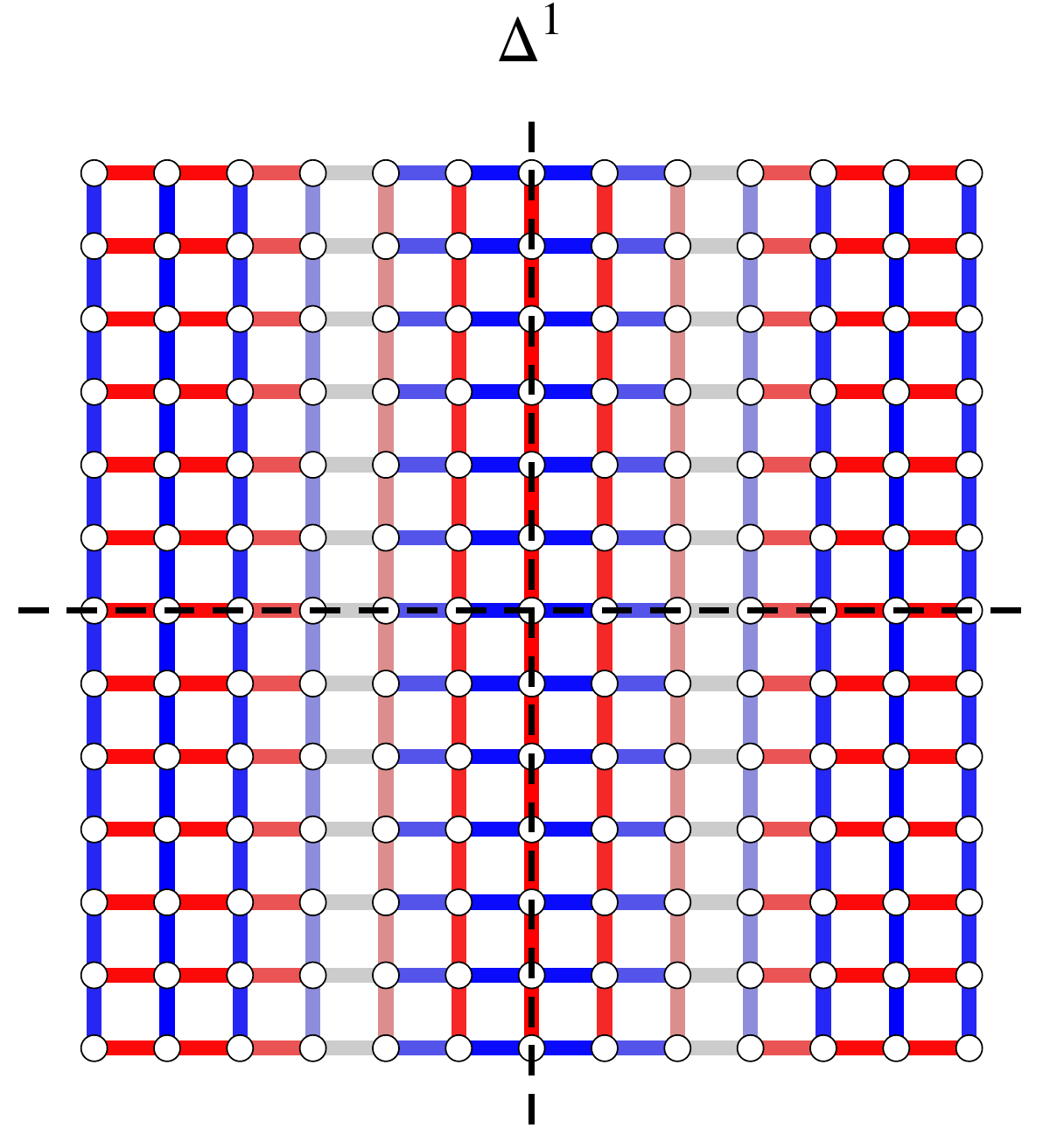}%
\includegraphics[scale=0.5]{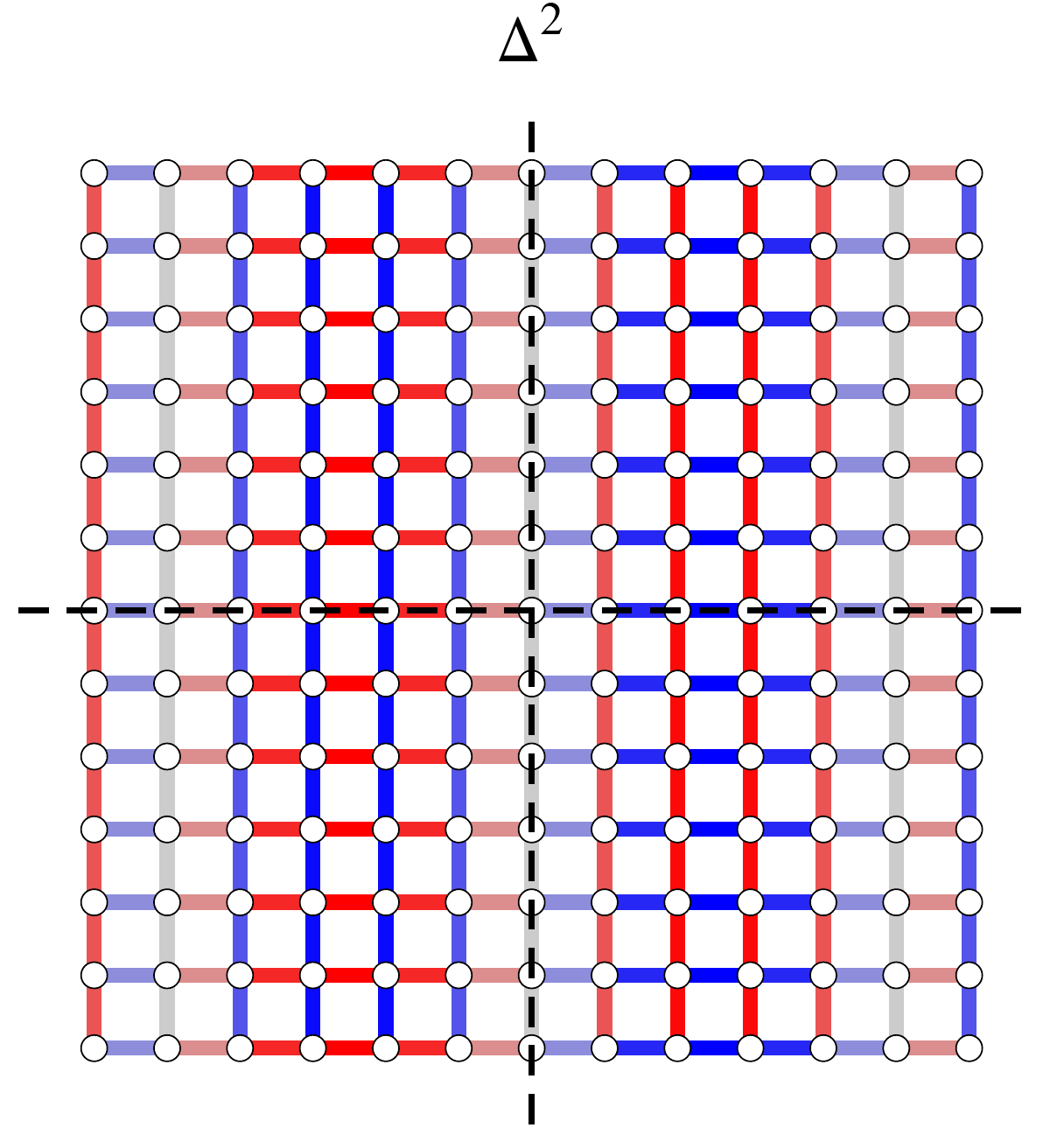}\includegraphics[scale=0.5]{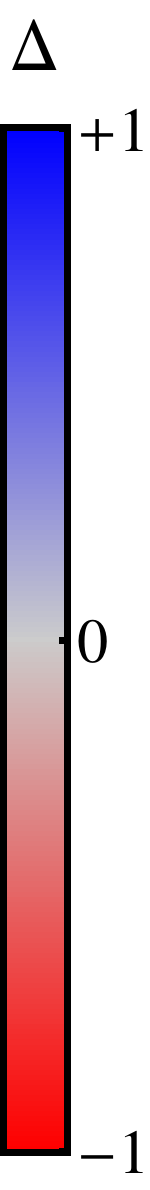}\\
\includegraphics[scale=0.5]{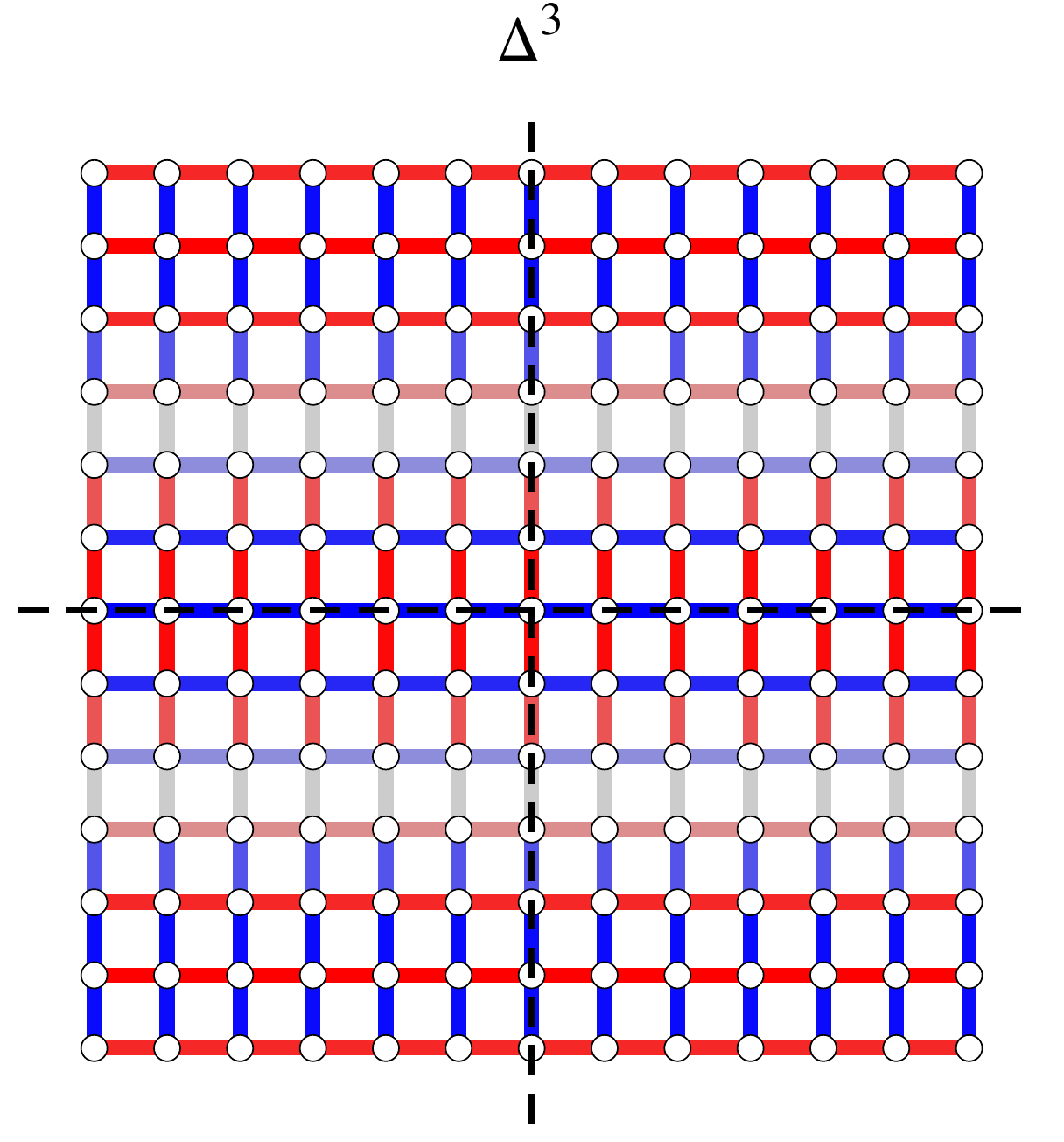}%
\includegraphics[scale=0.5]{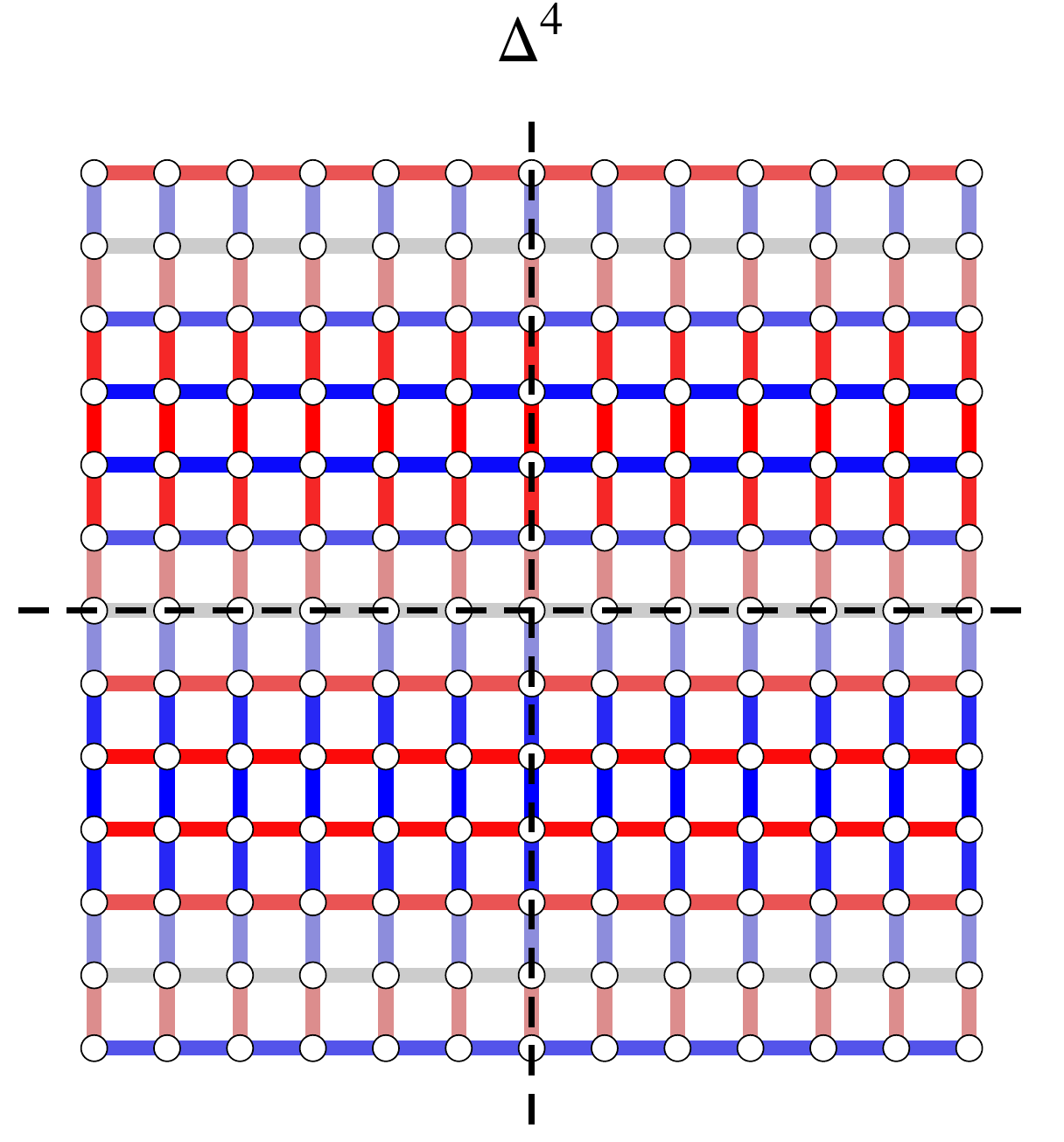}\includegraphics[scale=0.5]{legend.pdf}
\end{center}
\caption{\label{fig:basis}Basis functions with wavevector parallel to the axis, for $Q = \pi / 5$.}
\end{figure}

It is convenient to think about the problem in real space. Let us consider the following basis functions for the order parameter $\Delta_{\bvec r, \bvec r'} = \left\langle c^{\dag}_{\bvec r} c_{\bvec r'}\right\rangle$ :
\begin{align}
 &\Delta^1_{\bvec r, \bvec r+ \bvec{\hat x}} = \cos Q\, \left(\bvec r \cdot \bvec{\hat x} + \frac{1}{2}\right)\,, && \Delta^1_{\bvec r, \bvec r+ \bvec{\hat y}} = -\cos Q\, \bvec r \cdot \bvec{\hat x}\,, \\
 &\Delta^2_{\bvec r, \bvec r+ \bvec{\hat x}} = \sin Q\, \left(\bvec r \cdot \bvec{\hat x} + \frac{1}{2}\right)\,, && \Delta^2_{\bvec r, \bvec r+ \bvec{\hat y}} = -\sin Q\, \bvec r \cdot \bvec{\hat x}\,, \\
 &\Delta^3_{\bvec r, \bvec r+ \bvec{\hat x}} = \cos Q\, \bvec r \cdot \bvec{\hat y}\,, && \Delta^3_{\bvec r, \bvec r+ \bvec{\hat y}} = -\cos Q\, \left(\bvec r \cdot \bvec{\hat y} + \frac{1}{2}\right)\,, \\
 &\Delta^4_{\bvec r, \bvec r+ \bvec{\hat x}} = \sin Q\, \bvec r \cdot \bvec{\hat y}\,, && \Delta^4_{\bvec r, \bvec r+ \bvec{\hat y}} = -\sin Q\, \left(\bvec r \cdot \bvec{\hat y} + \frac{1}{2}\right)\,, \,.
\end{align} 

These basis functions are illustrated in fig.~\ref{fig:basis}. They are all real and hence they do not break time reversal. They are unidirectional, $d$-form factor bond order waves. $\Delta^1$ and $\Delta^2$ have wavevector parallel to the $x$ axis and differ by a phase. In the same way, $\Delta^3$ and $\Delta^4$ have wavevector parallel to the $y$ axis and differ by a phase. These four functions support a reducible representation of the point group:
\begin{align}
 &P \Delta^1 = \Delta^1\,,&& P \Delta^2 = -\Delta^2\,, && P \Delta^3 = \Delta^3\,,&& P \Delta^4 = \Delta^4\,,\\
 &R \Delta^1 = -\Delta^3\,,&& R \Delta^2 = -\Delta^4\,,&& R \Delta^3 = -\Delta^1\,, && R \Delta^4 = \Delta^2\,.
\end{align} 

This representation is decomposed in irreducible representations as follows:
\begin{itemize}
\item $\{\Delta^2,\ \Delta^4\}$ is a basis for the faithful representation $\Gamma_{\mathrm f}$.
\item $\{\Delta^1 + \Delta^3\}$ is a basis for $\Gamma_{\mathrm r}$.
\item $\{\Delta^1 - \Delta^3\}$ is a basis for the trivial representation $\Gamma_{\mathrm 0}$.
\end{itemize}

\begin{figure}[ht]
\begin{center}
\includegraphics[scale=0.5]{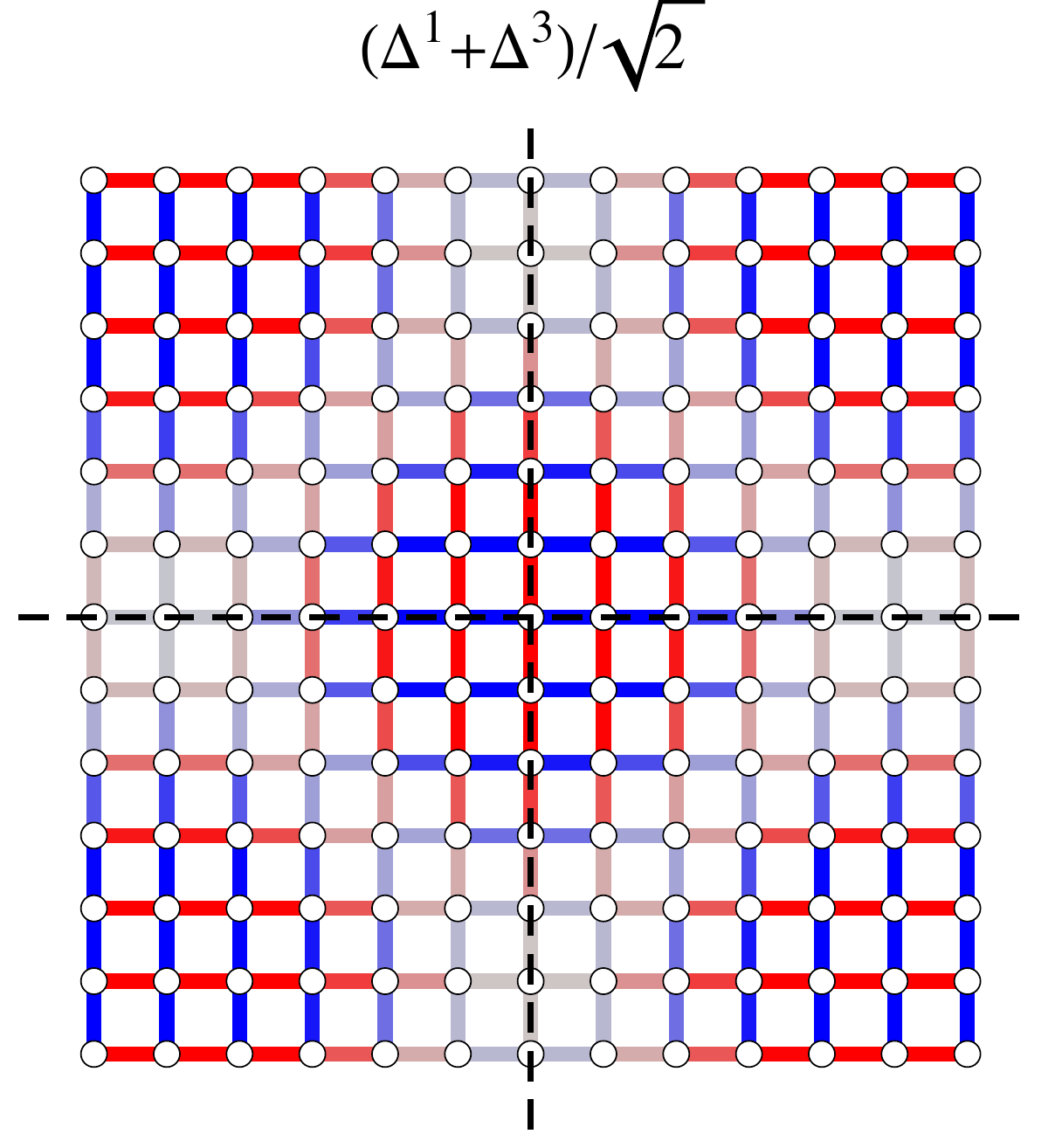}%
\includegraphics[scale=0.5]{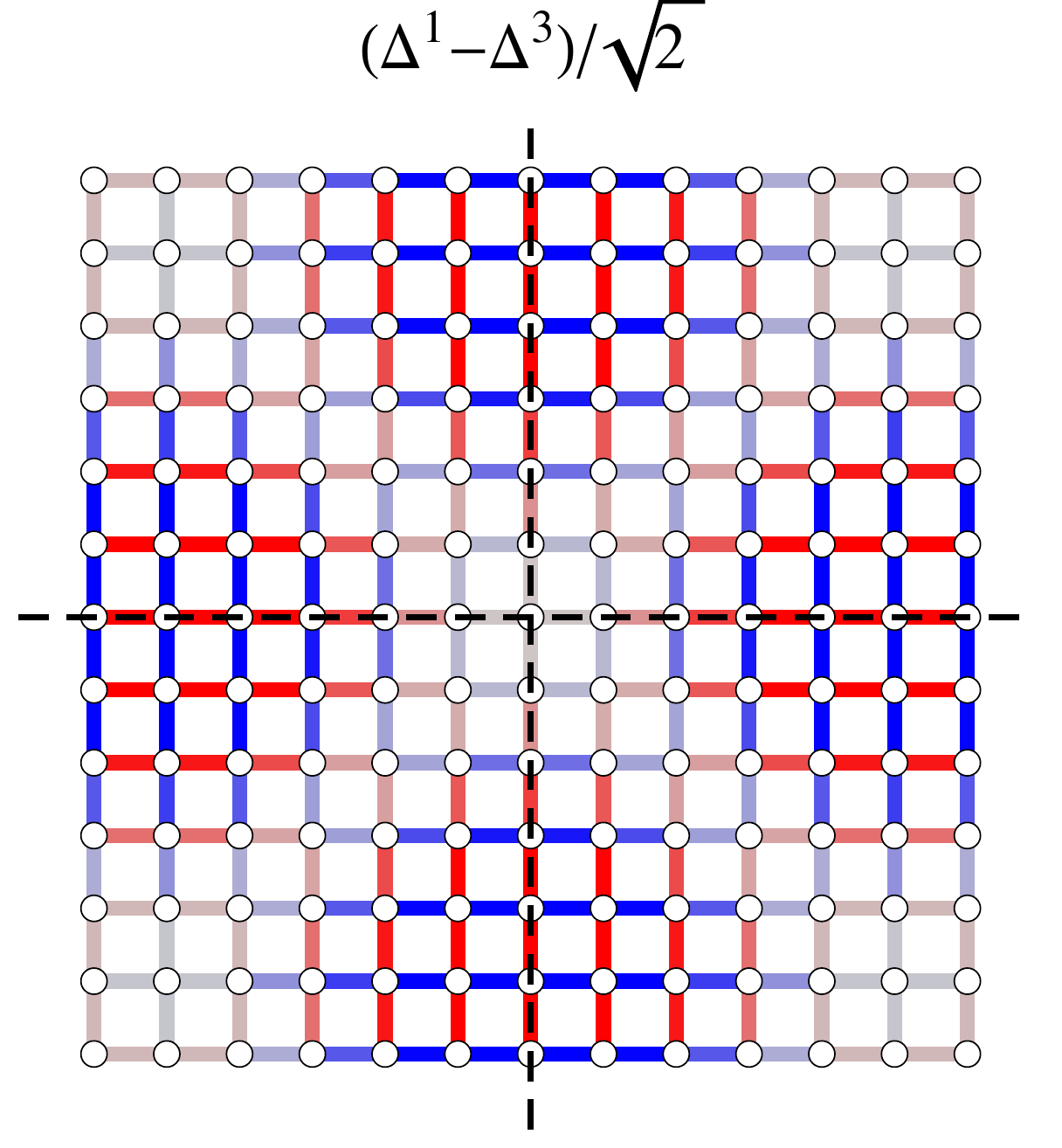}%
\includegraphics[scale=0.5]{legend.pdf}%
\end{center}
\caption{\label{fig:irred}Basis functions for the representation $\Gamma_{\mathrm r}$ (left) and $\Gamma_{0}$ (right), with wavevector parallel to the axis.}
\end{figure}

The two basis $\{\Delta^1 + \Delta^3\}$, $\{\Delta^1 - \Delta^3\}$ are shown in fig. \ref{fig:irred}.

\begin{figure}[ht]
\begin{center}
\includegraphics[scale=0.5]{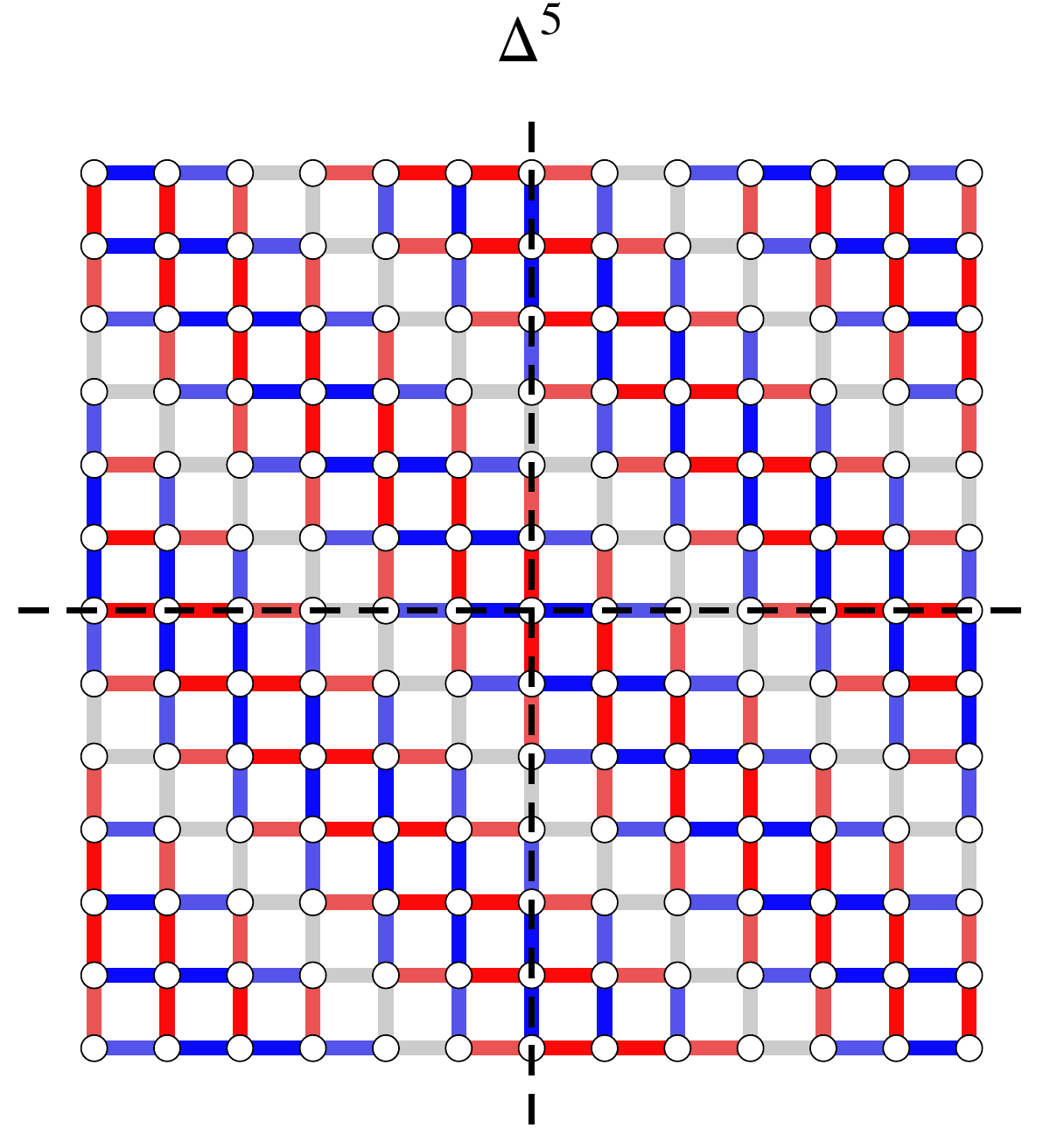}%
\includegraphics[scale=0.5]{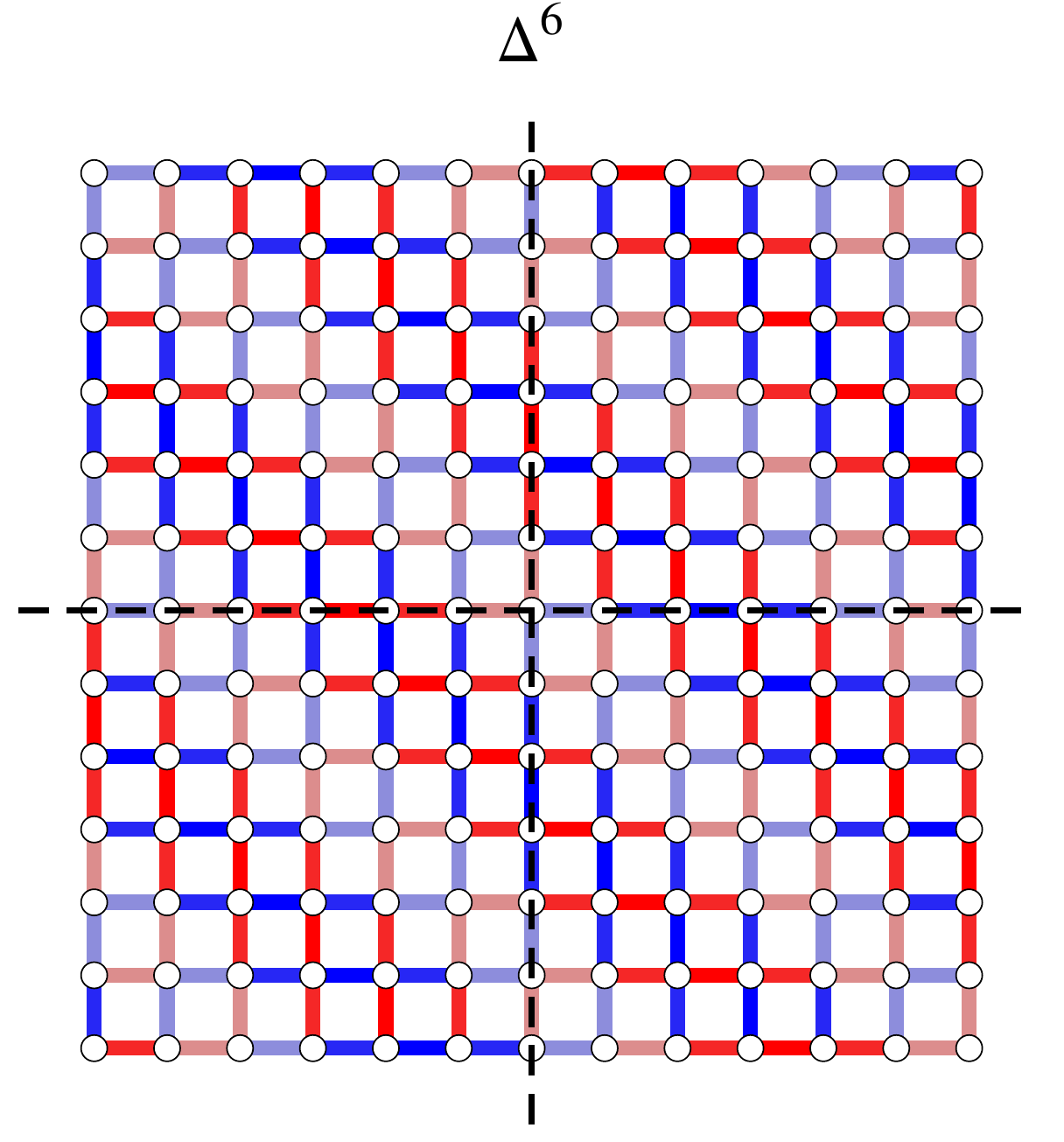}\includegraphics[scale=0.5]{legend.pdf}\\
\includegraphics[scale=0.5]{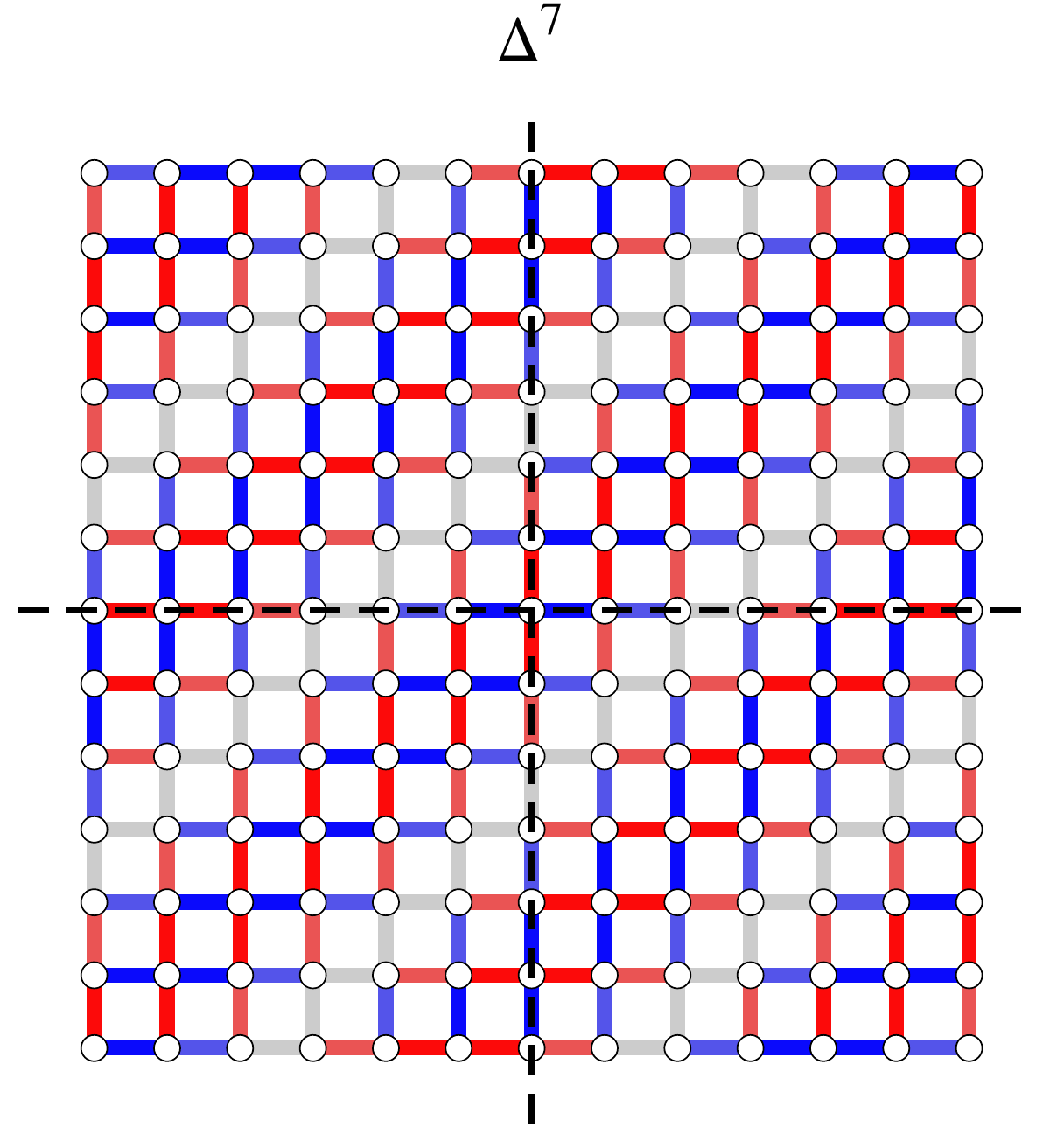}%
\includegraphics[scale=0.5]{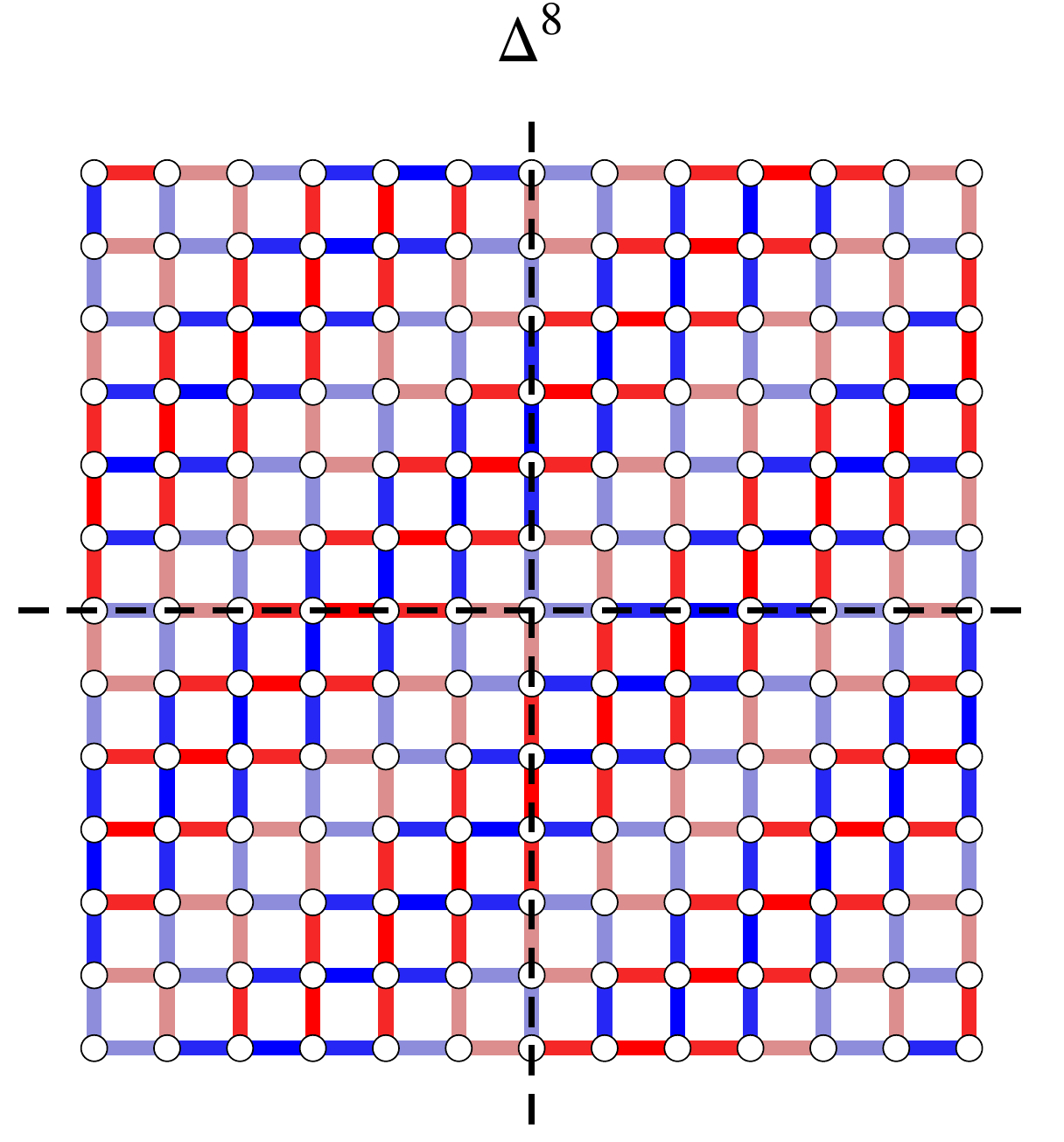}\includegraphics[scale=0.5]{legend.pdf}
\end{center}
\caption{\label{fig:basis_diag}Basis functions with diagonal wavevector, for $Q = \pi/5$.}
\end{figure}

The same procedure can be carried out for bond order waves with diagonal wavevector. Let us consider the following basis functions:
\begin{align}
 &\Delta^5_{\bvec r, \bvec r+ \bvec{\hat x}} = \cos Q\, \left(\bvec r \cdot \bvec{\hat x} + \bvec r \cdot \bvec{\hat y} + \frac{1}{2}\right)\,, && \Delta^5_{\bvec r, \bvec r+ \bvec{\hat y}} = -\cos Q\, \left(\bvec r \cdot \bvec{\hat x} + \bvec r \cdot \bvec{\hat y} + \frac{1}{2}\right)\,, \\
 &\Delta^6_{\bvec r, \bvec r+ \bvec{\hat x}} = \sin Q\, \left(\bvec r \cdot \bvec{\hat x} + \bvec r \cdot \bvec{\hat y} + \frac{1}{2}\right)\,, && \Delta^6_{\bvec r, \bvec r+ \bvec{\hat y}} = -\sin Q\, \left(\bvec r \cdot \bvec{\hat x} + \bvec r \cdot \bvec{\hat y} + \frac{1}{2}\right)\,, \\
 &\Delta^7_{\bvec r, \bvec r+ \bvec{\hat x}} = \cos Q\, \left(\bvec r \cdot \bvec{\hat x} - \bvec r \cdot \bvec{\hat y} + \frac{1}{2}\right)\,, && \Delta^7_{\bvec r, \bvec r+ \bvec{\hat y}} = -\cos Q\, \left(\bvec r \cdot \bvec{\hat x} - \bvec r \cdot \bvec{\hat y} - \frac{1}{2}\right)\,, \\
 &\Delta^8_{\bvec r, \bvec r+ \bvec{\hat x}} = \sin Q\, \left(\bvec r \cdot \bvec{\hat x} - \bvec r \cdot \bvec{\hat y} + \frac{1}{2}\right)\,, && \Delta^8_{\bvec r, \bvec r+ \bvec{\hat y}} = -\sin Q\, \left(\bvec r \cdot \bvec{\hat x} - \bvec r \cdot \bvec{\hat y} - \frac{1}{2}\right)\,, \,.
\end{align} 

These basis functions are illustrated in fig.~\ref{fig:basis_diag}. They are all real and hence they do not break time reversal. They are unidirectional, $d$-form factor bond order waves. $\Delta^5$ and $\Delta^6$ have wavevector parallel to $\hat x + \hat y$ and differ by a phase. In the same way, $\Delta^7$ and $\Delta^8$ have wavevector parallel to $\hat x - \hat y$ and differ by a phase. These four functions support a reducible representation of the point group:
\begin{align}
 P \Delta^5 = \Delta^7\,,&& P \Delta^6 = -\Delta^8\,,
\end{align}
\begin{align}
 R \Delta^5 = -\Delta^7\,,&& R \Delta^6 = \Delta^8\,,&& R \Delta^7 = -\Delta^5\,, && R \Delta^8 = -\Delta^6\,.
\end{align} 

This representation is decomposed in irreducible representations as follows:
\begin{itemize}
\item $\{\Delta^6,\ \Delta^8\}$ is a basis for the faithful representation $\Gamma_{\mathrm f}$. More precisely, the representation matrices have the form (\ref{faithful}) in the basis $\{\Delta^6 + \Delta^8,\, \Delta^6 - \Delta^8\}$.
\item $\{\Delta^5 + \Delta^7\}$ is a basis for $\Gamma_{\mathrm r}$.
\item $\{\Delta^5 - \Delta^7\}$ is a basis for the trivial representation $\Gamma_{\mathrm 0}$.
\end{itemize}

\begin{figure}[ht]
\begin{center}
\includegraphics[scale=0.5]{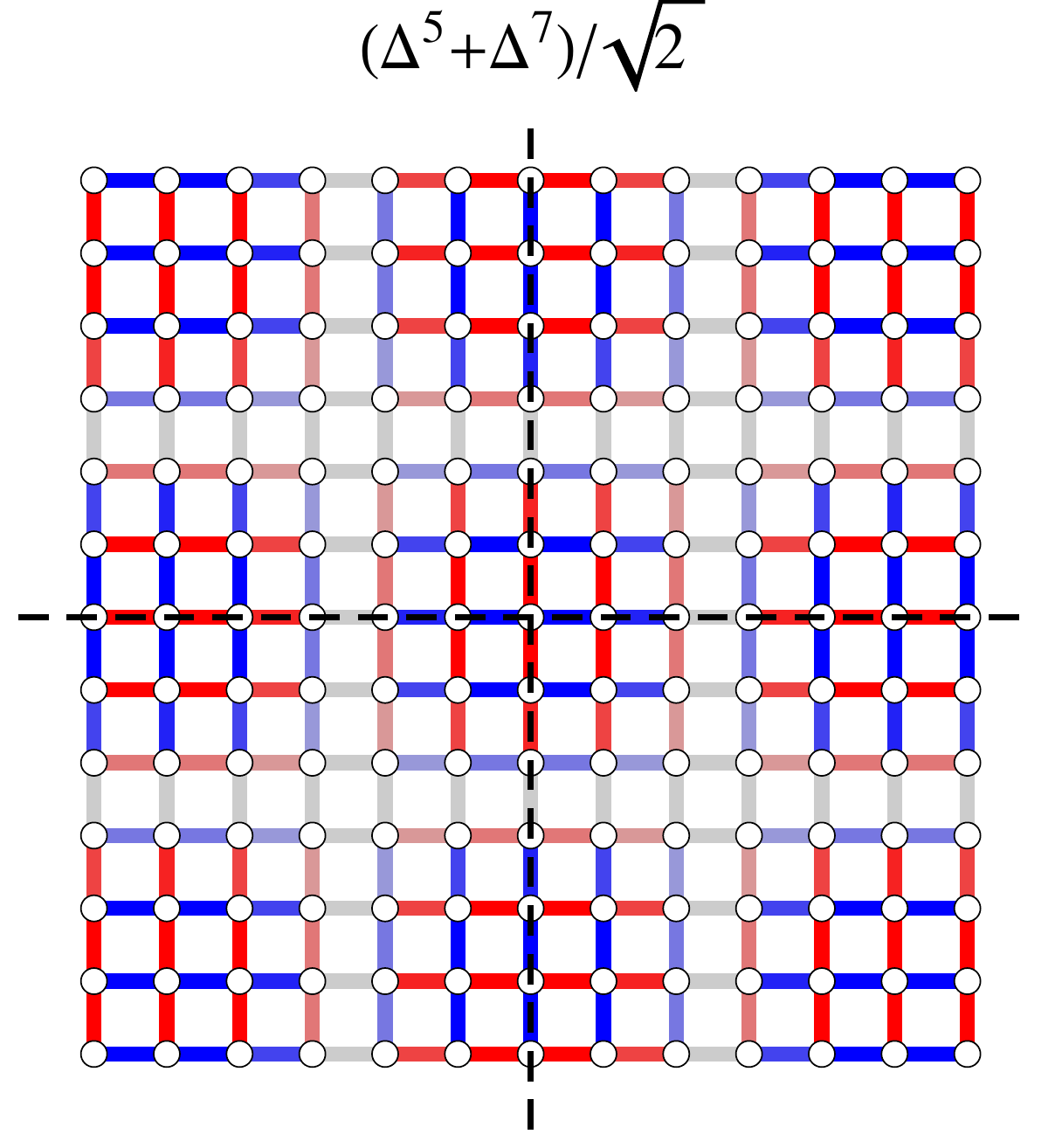}%
\includegraphics[scale=0.5]{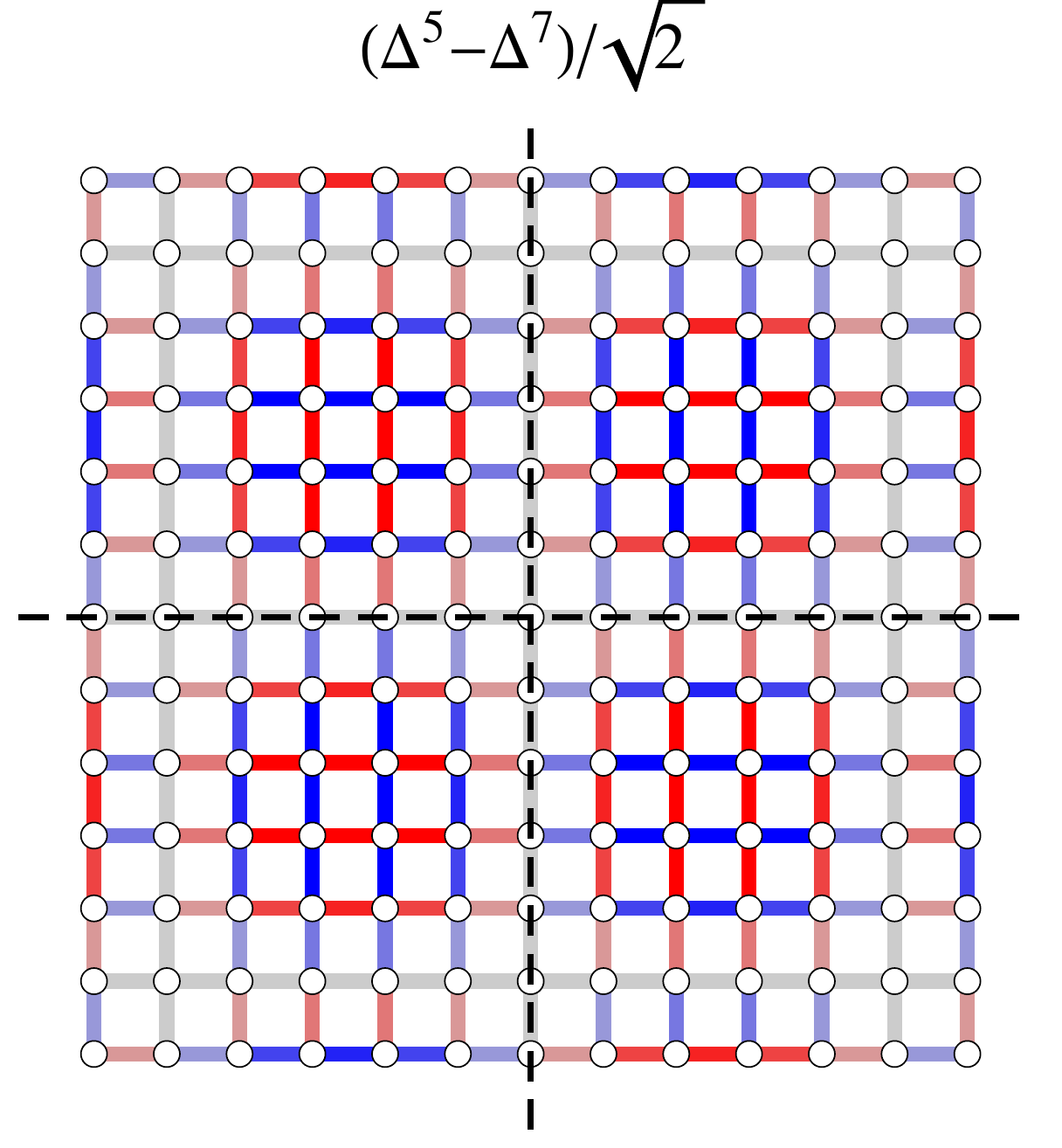}%
\includegraphics[scale=0.5]{legend.pdf}%
\end{center}
\caption{\label{fig:irred_diag}Basis functions for the representation $\Gamma_{\mathrm r}$ (left) and $\Gamma_{0}$ (right), with diagonal wavevector.}
\end{figure}

The two basis $\{\Delta^5 + \Delta^7\}$, $\{\Delta^5 - \Delta^7\}$ are shown in fig.~\ref{fig:irred_diag}. With reference to these two order parameters, the irreducible representation argument would tell us that the either the order $\Delta^5 + \Delta^7$ or $\Delta^5 - \Delta^7$ develops. However, by looking at fig.~\ref{fig:irred_diag}, it is clear that they are almost the same upon translation. They would be exactly the same if the wavevector had been chosen incommensurate. This means that the free energy splitting between the two irreducible representations is just a commensuration effect, and hence goes to zero as the commensuration period grows. This makes it clear that the irreducible representation argument is not of the greatest physical relevance in this context.

Zeyher discusses \cite{zeyher} how from possible degenerate ordering vectors $\bQ$ a
particular order develops. This depends on the parameters of an expansion
of the Landau free energy.
The Landau free energy only has $N$'th order terms of the form
\beq
\prod_{i=1}^N \Delta_{\bQ_i} (\bk_i) \label{prod}
\eeq
where translational symmetry requires that
\beq
\sum_{i=1}^N \bQ_i = 0
\eeq
for every term (up to reciprocal lattice vectors). 
So a state with incommensurate wavevectors $\bQ = (q,q)$ and $-\bQ$ will not necessarily 
generate density waves at $\bQ^\prime =(q,-q)$ because
there are no terms which are linear in $\Delta_{\bQ'} (\bk')$ in Eq.~(\ref{prod}).

\acknowledgments

We thank R. Zeyher for helpful discussions.
This research
was supported by the NSF under Grant DMR-1103860, the Templeton foundation, and MURI grant W911NF-14-1-0003 from ARO.
J. B. acknowledges financial
support from the DFG through grant number BA 4371/1-1.
Research at Perimeter Institute is supported by the Government of Canada through Industry Canada and by the Province of Ontario through the Ministry of Economic Development \& Innovation.

\end{document}